\documentstyle[preprint,aps]{revtex}
\newcommand{\be}{\begin{equation}}
\newcommand{\ee}{\end{equation}}
\newcommand{\slp}{\slash\hspace{-2.2mm}p}
\newcommand{\sln}{\slash\hspace{-2.2mm}n}
\newcommand{\slq}{\slash\hspace{-2.2mm}q}
\input epsf.sty
\begin{document}
\draft
\title{Photodisintegration of the deuteron in the few GeV region
 using asymptotic amplitudes}
\author{A.E.L.Dieperink$^a, \ $S.I.Nagorny$^b$    }
\address{ $^a$ \it  Kernfysisch Versneller Instituut, NL-9747 AA Groningen}
\address{$^b$ \it Nikhef, P.O.Box 41882, NL-1009 DB Amsterdam, The Netherlands}
\maketitle
\begin{abstract}
Exclusive photodisintegration of the deuteron in the 1-4GeV range is described in terms
of a simple covariant and gauge invariant approach using an effective counting rule
as the  hard part of the d-np vertex.
At a  scattering angle of $\theta_{cm}=90^\circ$  a prescaling behavior of the
differential cross section $\propto s^{-(n-2)}$ with $n \approx 12 $ is obtained;
 going away from $90^\circ$ the value of $n$ decreases slowly, in qualitative agreement
with the recent data.
 \end{abstract}
\section{Introduction}
The exclusive process of photodisintegration of the deuteron
addresses an interesting interplay of nuclear and particle physics.
At low energies (say, below $E_\gamma$ = 0.5 GeV) conventional nuclear
 models based upon meson exchange which fit the NN phase shifts
 give a satisfactory description of both the energy dependence and the angular
distribution of the experimental cross section
\cite{Lee,Aren,Laget}. However,
at higher energies nuclear potential models fail to explain the data
\cite{Belz,Bochna}.
This is not unexpected since at $E_\gamma >1$ GeV small distances of the
order of 0.2 fm play
a role. \\ Alternatively it has been attempted
 to describe the cross section in terms of quark-gluon degrees of freedom \cite{BF,BH,Kon}.
   A possible signature for the emergence of
quark-gluon degrees of freedom would be
 the observation of the onset of scaling of the cross section
\cite{BF}. Several examples where this happens have been
found. For example, the $p(\gamma ,\pi) n$ reaction above 3 GeV
appears to be consistent with counting rules at all angles {\cite{And}. \\
In case of the exclusive process $d(\gamma,p)n$
the dimensional counting analysis of \cite{BF} leads to a differential cross section
of the form
\be
     d\sigma/dt = s^{-(n-2)} f(\theta),
\ee
where $s (\theta)$ is the cm energy (angle),
and $n$ denotes the number of elementary fields in initial and final state
(i.e. $n=13$ in case of the deuteron).
\\ Previous data from SLAC \cite{Belz} up to $E_\gamma$ =3 GeV at
$\theta =90^\circ$  indicated a scaling behavior  consistent with $n=13.1 \pm 0.3.$
More recent data from Jefferson Lab up to 4GeV confirmed \cite{Bochna}
the scaling behavior with $n \approx 13$  at $\theta=90^\circ$ and $69^\circ; $
 whereas at smaller angles, $\theta= 36^\circ , 52^\circ, $
 the best fit yielded $n \approx$ 11.5 and 11.6, respectively.
 This constitutes a deviation from the scaling behavior predicted by simple counting rules.
\\ A more refined approach is
the reduced nuclear amplitude (RNA) approach  \cite{BH} which is also based upon
parton exchange between the two nucleons, but  takes into account some
finite mass and higher twist effects. However, if normalized
at $E_\gamma =1.0 $GeV,
the prediction falls below the data at $\theta=90^\circ $ for $E_\gamma > 3$GeV.
On the other hand
the quark-gluon string (QGS) model proposed in \cite{Kon} and based upon Regge
phenomenology appears to describe the data \cite{Bochna}
only at small $t$ values, corresponding to small angles. \\
This indicates that in practice the situation is more complex.
\\ The aim of this paper is to study the question of the possible origin of
the  apparent scaling
and scaling violation  in more detail; in particular we address
whether the occurrence of scaling is an exclusive
pQCD phenomenom or whether it can arise from different mechanisms.
\\ The  approach in the present paper, in which the basic degrees of freedom
are taken to be hadronic, is an extension of the one in ref. \cite{Nagorny}
 which predicted
 the cross section at $\theta=90^\circ$ for $E_\gamma >1$ GeV in
fair agreement with experiment, but failed to describe the angle dependence.
Using a simple covariant parametrization of the deuteron vertex in terms of
 a hard component (suggested by an effective counting rule) and
imposing gauge invariance  the cross section for large
but not infinite $s$ shows a ``preasymptotic" scaling
behavior at $\theta=90^\circ$. This resembles the counting rule prediction,
however, with an angle dependent value of $n$ (which decreases away from
$90^\circ$).
\section{Formalism}
The general covariant half off-shell d-pn vertex ($p^2_2=m^2$)
can be expressed as \cite{SB}
\be
  A^\mu= \Gamma_1 \gamma^\mu +\frac{(p_1-p_2)^\mu}{2m} \Gamma_2 +\frac{\slp_1-m}
  {2m}[\Gamma_3\gamma^\mu + \frac{(p_1-p_2)^\mu }{2m}\Gamma_4],
   \label{Avertex}
\ee
  where the $\Gamma_i$ are scalar functions: invariant form factors. For
large virtualities
the contribution of the first term in eq.(\ref{Avertex}), proportional to
$\Gamma_1, $ dominates over the last three. \\
Assuming that at large $s$ only tree-type diagrams survive,
the $ \gamma +d \rightarrow n+p$ (Born) amplitude can be written as  the
sum of the pole diagrams in the $s,t$ and $u$ channels ($\Gamma_1 \equiv G$):
$$
   T^{\mu}_{pole}= \xi_{\nu}(d) \bar{u}(p')
      [F^p_\mu
     \frac{ \slp +m}{ p^2-m^2} \gamma^\nu G(-k_t^2)
    +  F^n_\mu
     \gamma^\nu \frac{ \sln-m}{n^2-m^2}
     G(-k_u^2)
 $$
\be + G(-k_s^2)\gamma^\alpha  \frac{ -g_{\beta \alpha}
       + d'_\beta d'_\alpha /d'^2}{ s-m_d^2} F^d_{\mu\nu\beta}] C \bar{u}^T(n').
\label{Tpole}
\ee
Here $s$-,$t$-, $u$- variables for the d-pn vertex are:
$k_t= (k_s-q)/2, \ k_u=(k_s+q)/2 $, \ $k_s=(p'-n')/2, \ p'=p+q, \
d'=d+q; $
$u(p)$ is the   nucleon spinor, 
 $\xi_{\nu}(d)$ the polarization vector of the deuteron
($\xi^{\nu}(d) d_{\nu} = 0$), $C $ is the charge conjugation operator,
$F_\mu^i \ (i=n,p) $ denotes the electromagnetic coupling to
the nucleon,
$ F^i_\mu= (e_i+ \slq \frac{ K_i}{2m}) \gamma^\mu $,
and $ F^d_{\mu\nu\beta}$ the corresponding one for the deuteron \cite{Nagorny}
($f_d = 2\mu_d-1+Q_d$):
$$-F^d_{\mu\alpha\beta}= 2d_\mu(g_{\alpha\beta} - q_\alpha q_\beta {f_d \over
2m_d^2})
  +2 \mu_d(g_{\mu\alpha} q_\beta- g_{\mu \beta} q_\alpha)
   + (s-m_d^2)( g_{\mu\alpha} q_\beta+g_{\mu \beta} q_\alpha) {f_d \over
4m_d^2}.$$
Both $\gamma NN$ and $\gamma dd$ vertices satisfy the identities:
$q^{\mu} F^i_{\mu} = e_i \slq$, $q^{\mu} F^d_{\mu \alpha \beta} =
-(s - m_d^2) g_{\alpha \beta}$.
Note, $s$-channel accounts for the pole
part of the $T$-matrix
of the final state n-p interaction.
\\ In the presence of a momentum dependent vertex function
the pole diagrams themselves are not gauge invariant. Indeed, using
 the Ward-Takahashi identity for the 3-point electromagnetic
vertices, eq.(\ref{Tpole}), and the Dirac equation one finds that the contraction of $q_\mu$ with
 the sum of the $s$-, $t$-, $u$-pole amplitudes does not vanish
(i.e. the corresponding Born current is not conserved), if the strong d-pn
vertices contain momentum dependent form factors
$$q_{\mu} T^{\mu}_{pole} = - \xi_{\nu}(d) \bar{u}(p') \gamma^{\nu} [ G(-k^2_t)
- G(-k^2_s) ] C \bar{u}^T(n') \neq 0.$$
Therefore the Born current in eq.(\ref{Tpole}) is not complete and a
{\it contact}
contribution (which should not contain any pole-type singularities!)
must be added to provide current conservation on the tree-level.
To this end we use {\it minimal insertion} of the gauge field directly
into the d-pn vertex \cite {Nag89}, which gives rise to a {\it contact
amplitude}, $T_c^{\mu}$:
\be
 T^{\mu}_c= \xi_{\nu}(d) \bar{u}(p') \gamma^\nu
   \int_0^1 \frac{d\lambda}{\lambda} \frac{ \partial}
{\partial q_\mu} \{e_p G(-(k_s-\lambda q/2)^2) +
   e_n  G(-(k_s+\lambda q/2)^2)\} C \bar{u}^T(n') . \label{Tc}
 \ee
Calculating the contraction $q_{\mu} T_c^{\mu}$ with $e_p=1$ and $e_n=0$,
one can easily check that the total current is conserved \cite{Nagorny}:
$  q_\mu (T^{\mu}_{pole}+ T^{\mu}_c) =0, $ irrespective of the
explicit form of the strong form factor $G(-k^2)$, and hence
 the total amplitude is gauge invariant.
\\ To proceed we need make a choice for the vertex function $G(-k_i^2).$
  Previous work \cite{Lee} showed that in the energy region above
$E_\gamma >$ 1GeV  a conventional potential model
wave function cannot describe the data.
\\ In the present approach  it is hypothesized that the d-np vertex
  can be separated into two  parts: a {\it soft} part corresponding
to conventional meson exchange theory, and
a small {\it hard}  component caused by short-range phenomena.
It is assumed that the soft part describes all low-energy
(static) properties of the NN system and provides the dominant contribution
to the normalization of the bound state wave function, while the hard part
 dominates the cross section at large virtualities.
Since the microscopic structure of the short-range
dynamics is poorly known we will use an effective counting rule prescription
\cite{Gross} to describe the hard part of the d-pn vertex:
\be
 G(p^2)=    \frac{C}{(\Lambda^2/2+m^2-p^2)^g }, \label{vertex}
\ee
where $p$ is the momentum of the off-shell nucleon,
$C$ is a normalization parameter and $\Lambda$ is related to the inverse of
the range.
For the special case $g=3$ \cite{Gross}
the three-pole vertex
represents one meson propagator and two (monopole) nucleon-meson form factors.
 \\
At the large virtualities involved obviously relativistic effects play an
important role. In practice there exist various relativistic formulations,
such as the instant form (if})
formalism and the light-front (lf) approach. Whereas in
an exact calculation these are expected to yield the same result, this is not
true in a truncated Fock space scheme. Indeed it has been noted \cite{Coes,FS}
that in lowest order (IA) the  lf and  if approaches lead to
different results. In particular
in the lowest order Fock states in the lf approach
 negative energy states do not enter.
To illustrate this model dependence
in the following we distinguish the covariant (instant form) and the
light-front  approach.
\\ {\bf Covariant Approach: Instant form kinematics}
\\ In terms of the variables $k_i$ the vertex in eq.(\ref{vertex})
 takes on the form ($i=s,t,u)$
\be
   G^{\rm{if}}(-k_i^2)= \frac{C}{ (\delta^2-k_i^2)^g}, \label{Gif}
\ee
   where $k^2_t=-\alpha \vec{k}^2$, $k_u^2=-(2-\alpha) \vec{k}^2, $
$ k_s^2=-\vec{k}^2$, with $\vec{k}^2 =s/4-m^2$ and
$\alpha=2pq/(dq)= 1-\frac{k}{E_k} \cos \theta. $
Furthermore $\delta^2= \Lambda^2/4+ \alpha_0^2 $
with $\alpha_0^2 = m^2- m_d^2/4.$
\\ Substituting
eq.(\ref{Gif}) in eq.(\ref{Tpole}) the cross section can be obtained
straightforwardly.
  The absolute value of the cross section cannot be
predicted and therefore the (only) parameter $C$ is
fixed by fitting the data at $E_\gamma$=1 GeV.
We have chosen this energy for the normalization since it is
in this region that
the microscopic d-np vertex gives a reasonable description of the absolute value
of the empirical cross section. \\
The resulting cross section $s^{11} \frac {d\sigma}{dt} $ at $\theta=90^\circ$
is compared with data in fig. 1 for  $g=3$ and  $g=4$.
For comparison the results for the reduced
amplitude approach of ref \cite{BH}
 which is very close to that $g=4$,
are also shown. It is seen that
 the observed overall energy dependence of the cross section
in the  energy region $1  < E_\gamma < $  4GeV is described well with $g=3$.
Only at the highest energies a larger value of $g$ would fit better.
\\ Turning to the angular distribution one sees from fig. 2 (dot-dashed curve)
that
 in the if approach the predicted angular dependence on $\theta$  increases
 rapidly away from $ 90^\circ.$  Although the data are available only
for a few angles this is clearly  not observed experimentally.
\\ {\bf Light-front approach} \\
Here the light-cone variables $\alpha$ and $k_\perp$ have a direct physical
interpretation as the longitudinal and transverse momentum fraction
carried by the nucleon in the deuteron, with
$ \vec{k}^2=( s-4m^2)/4=  \frac{m^2(1-\alpha)^2+ k_{\perp}^2}{
\alpha(2-\alpha)}. $
We will oriente the normal vector of the light-cone hypersurface along $\vec{q}$
to suppress $Z$-graphs \cite{Nagorny}.
 In general strong lf form factors are functions of two
variables, for which we can use any convenient pair from the set
$(\alpha,k_\perp,k_3, \vec{k}^2)$.
For simplicity we will assume a factorized form of the  lf form factor:
 \be
  G^{\rm lc}(\alpha, \vec{k}^2)=   \frac{C} {(
    \beta^2-\vec{k}^2)^g} \phi_\kappa(\alpha),  {\rm \ with \ } \phi_\kappa
  (\alpha)= \alpha^{-\kappa}.  \label{Glc}
 \ee
Here the functional dependence on $\vec{k}^2$ is the same as in the instant form,
 eq.(\ref{Gif}).
The function $\phi(\alpha)$  in eq.(\ref{Glc})
goes to unity in the nonrelativistic limit.
 The simplest choice  is $\kappa=0,$
(as in ref \cite{FS}), 
but a more realistic choice is $\kappa=\frac{1}{2} $ \cite{Kar}
(basically corresponding to the Wick Cutkovsky model).
At $\theta= 90^\circ $ (where $\alpha=1$) both vertices
 are identical and
the  lf and  if formalisms lead to the same cross cross section
 (apart from a slight difference coming from the contact term, eq.(\ref{Tc})).
However, at $\theta \neq 90^\circ$ the form factors (6) and (7) reproduce
a completely different dynamics.
Note, that the angular dependence of the cross section arises
mainly from the
dependence of the arguments of $G$ in eq.(\ref{Tpole}) on
 $\alpha= (1 - \frac{k}{E_k} \cos \theta). $
\\ In  $G^{\rm if}$  in eq.(\ref{Gif})
these arguments  are    $k_t^2= -\alpha \vec{k}^2$ and $k_u^2= -(2-\alpha)
 \vec{k}^2 $. This gives rise to
 a  strong increase of the cross section at forward and backward angles,
which has a dynamic origine. Namely because of the requirement of covariance
 in the instant form the vertex dependence on
$\alpha$  and $\vec{k}^2$  effectively reduces to the dependence on one
covariant variable only, say  $k_t^2$ (or $k_u^2$); the latter
 is a function of both $s$ and $\theta$, and hence in this case
the angle dependence  is essentially dictated by the value of the
vertex parameter $g$ in eq.(\ref{vertex}),
which also determines the $s$- dependence.
\\ On the other hand in the  lf approach the angular distribution
 is flatter since in this case the vertex depends on two variables,
$\alpha$ and $\vec{k}^2,$ which  are in principle  independent.
Thus, a steep  $s$-dependence of the cross section may be
consistent with a flat $\theta$-dependence.
 One sees from the  curves in fig.2 that with $\kappa =\frac{1}{2}$
 the observed angular dependence at $E=3.2$ GeV is described well.
\\ In fig.3 the calculated energy dependence of the cross sections at
 4 different angles is compared with the data.
 While the
overall agreement is reasonable  the data at the  forward angles
suggest a steeper increase of the cross section with energy than predicted,
and moreover there is a discrepancy for the highest energy at $69^\circ.$
Also shown are
 the results of the RNA  \cite{BH} and (except at $69^\circ$) the QGS approach \cite{Kon};
the latter is expected  to be applicable only at small angles.
\section{Discussion}
It is of interest to discuss the underlying mechanism for scaling in the present approach.
Using eqs.(\ref{Avertex}-\ref{Tc})
 we can express the cross section for large (but finite) $s$
\be
\frac{d\sigma}{dt}= \frac{m_dC}
 {\sqrt{s}(s-m_d^2)^{2g-1}(s-m_d^2)^{3/2}} f(\theta,s)  \label{as}.
\ee
For $\theta =90^\circ$ one has $\alpha=1$ and  $f(\theta,s)=1$.
 In the relevant region of $s$ the contribution of $m_d^2$
in eq.(\ref{as}) is still non-negligible;
 for $g=3$ the calculated ``prescaling" behavior in the relevant energy region
can be approximately written as  $\frac{d\sigma}{dt} \approx s^{-n+2},$
with $n-2=10. $ 
\\ For other angles
 $f(\theta,s)$ depends upon $s$  through $\alpha(\theta,s)$
and therefore is a different function of $s$ at different angles, effectively
giving an additional power of $s$ in eq.(8) when $\theta \neq 90^\circ$.
 Therefore only for $s \rightarrow \infty $ the longitudinal fractions $\alpha$
and $\alpha' = 2-\alpha$ do not depend upon $s$: $\alpha(s \rightarrow \infty)
\rightarrow 1- \cos(\theta)$, and one expects an (angular independent)
scaling. We find that in the present model in the energy region
1-4 GeV $n$ decreases slowly from $n-2 \approx 10 $ at
$90^\circ$ to $n-2 \approx 8 $ at $10^\circ$ (or $170^\circ$).
In practice from eqs.(\ref{Avertex}-\ref{Tc})
this behavior can be expressed as an angular dependence of the cross section on
the  lf variable $\alpha$, namely $\frac{d\sigma}{dt} \propto \alpha^{-2}
\ (\alpha >1)$ or $\propto \ (1-\alpha)^{-2} \ (\alpha < 1). $
\\ In the past Brodsky \cite{BH} has discussed the concept of
a ``hadron helicity conservation law".
 In case of the $\gamma+d \rightarrow p+n$ reaction it states that
only helicity amplitudes satisfying the
condition $\lambda_p + \lambda_n = \lambda_d$
contribute at $s \gg m^2$.
Taking into account that only the Dirac coupling ($ \approx \gamma_{\mu}$)
in the $\gamma NN$-vertex conserves  hadron helicity,
while Pauli coupling ($\approx \sigma_{\mu \nu}$) does not,
 we see that in the present approach  asymptotic ``hadron
helicity conservation"
will occur only in  case the Pauli couplings are negligible in the limit
$s \gg m^2$, i.e. if the latter
 fall off at least as $1/s^2.$  
 Note that the gauge constraint for the 3-point EM Green function,
 which does not allow any form factors in the Dirac coupling
 (in case of a  reducible vertex with real photon),
 does not lead to any restrictions for the Pauli one.
\\ In practice helicity conservation predicts that for $s \gg m^2$  the
cross section in the backward hemisphere receives no contribution
from the neutron pole (located near
 180$^\circ)$ but only from
  the deuteron pole which does not depend on the angle.
 Hence at backward angles the cross section would scale with $s$
independent of angle.
 On the other hand at forward angles, where we have a competition of  two
different pole contributions, namely the proton one
(which depends on $\theta$ and $s$) and the deuteron one,
one does not expect a unified $s$-dependence
but rather an angle dependent  scaling.
For this reason it would be of interest to extend the experiment to
backward angles.\\
As to the  differences in the  results in the lf and if formalisms
we note that we used an effective counting rule for the hard
part of the d-pn vertex; this is based on the assumption
of a {\it fixed number of constituents} which is only well defined
in the lf approach, but not in the if. Therefore we consider
the results of the latter (in this particular model)  less reliable
at extreme angles, which involve large $t$-, $u$-virtualities.
\\  We note that the deuteron vertex at large virtualities
  can in principle  be addressed in more detail in semi-exclusive reactions,
 such as
$ d(e,e'p)X$ at large $Q^2$ in which the spectator proton is observed in
 the backward hemisphere to avoid contamination from hadronization products.
 In the past this reaction has been proposed
\cite{Melnit} to discriminate between
 various models for the nuclear EMC effect. In the IA the deuteron  hadronic
tensor factorizes in terms of a neutron structure function
 and a nuclear spectral function determining the variables $\alpha, k_\perp$
of the observed recoiling proton. Hence the cross section is directly
proportional to the square of the  deuteron vertex.
In this respect we note that the relation between the spectral function
 and the deuteron wave function is different for the lf and if
formalism \cite{Melnit,FS}.
(In fact in the covariant formulation it is not well possible to
define a proper normalization.)
 This difference  tends
to increase with increasing $\alpha$, and depend also
on $k^2_\perp$ \cite{Lex}.
\\ Finally we note that
in this work we have assumed (as an extreme model) that
  the hard part of the NN interaction resides only in the (initial) deuteron
vertex, and
not in the final np state; in future work we will explore the contribution from
hard contributions in the final np rescattering. \\
 {\bf Acknowledgement \\}
The authors are indebted to R.J. Holt, A. Radushkin and B.P.Terburg
 for stimulating discussions.
This work is supported in part by the Stichting voor Fundamenteel
Onderzoek der Materie (FOM) with financial support from the
Nederlandse Organisatie voor Wetenschappelijk Onderzoek (NWO).
\\  {\bf Figure Captions} \\
Fig. 1. Calculated energy dependence of $\frac{d\sigma}{dt}s^{11}$ at $\theta=90^
\circ $ for $g=3$ (dashed), 4 (dotted curve);
the solid curve is the result from
Brodsky and Hiller \cite{BH}; the data are from \cite{Belz,Bochna}.
\\ Fig. 2. Calculated angular dependence of $s^{11} \frac{d\sigma}{dt}$
 at $E_\gamma= 3.2$ GeV , for $g=3,$ in the light-front approach
 with $\kappa $=0 (solid),$\frac{1}{2}$
 (dotted), 1 (dashed), and instant form (dashed-dotted curve). \\
Fig. 3. The scaled cross section $s^{11}\frac{d\sigma}{dt}$ as a function
of photon energy for (a) $\theta =89^\circ $, (b) $69^\circ, $
(c) $52^\circ$, (d) $ 36^\circ$.
The data are from ref. \cite{Belz,Bochna},
 the solid curve shows the present results, the dashed curve those from
\cite{Kon} (not available for $69^\circ),$ and the dotted curve those
from \cite{BH}.
\newpage
\epsfxsize=20cm
\epsfbox{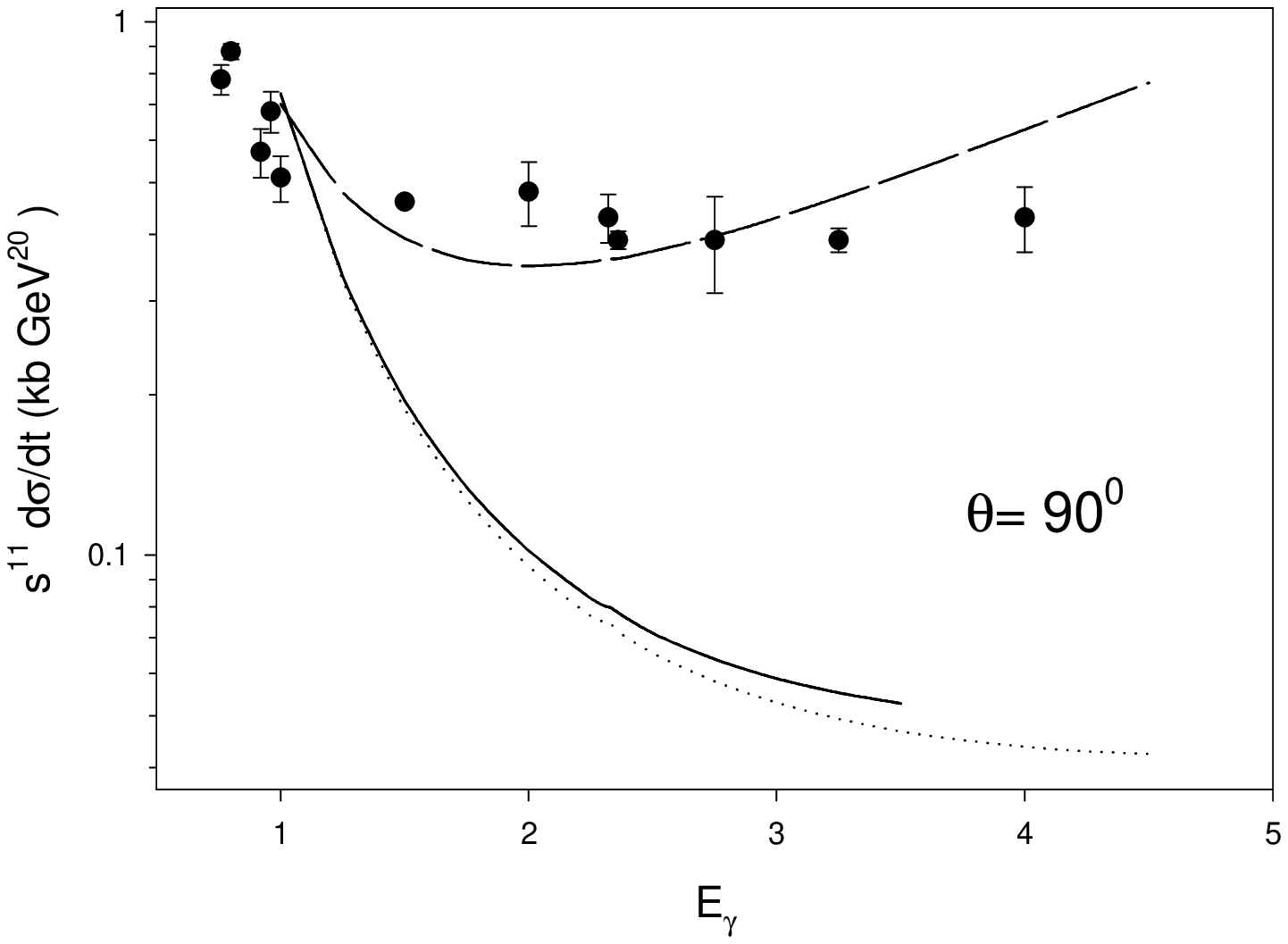}
\newpage
\epsfxsize=20cm
\epsfbox{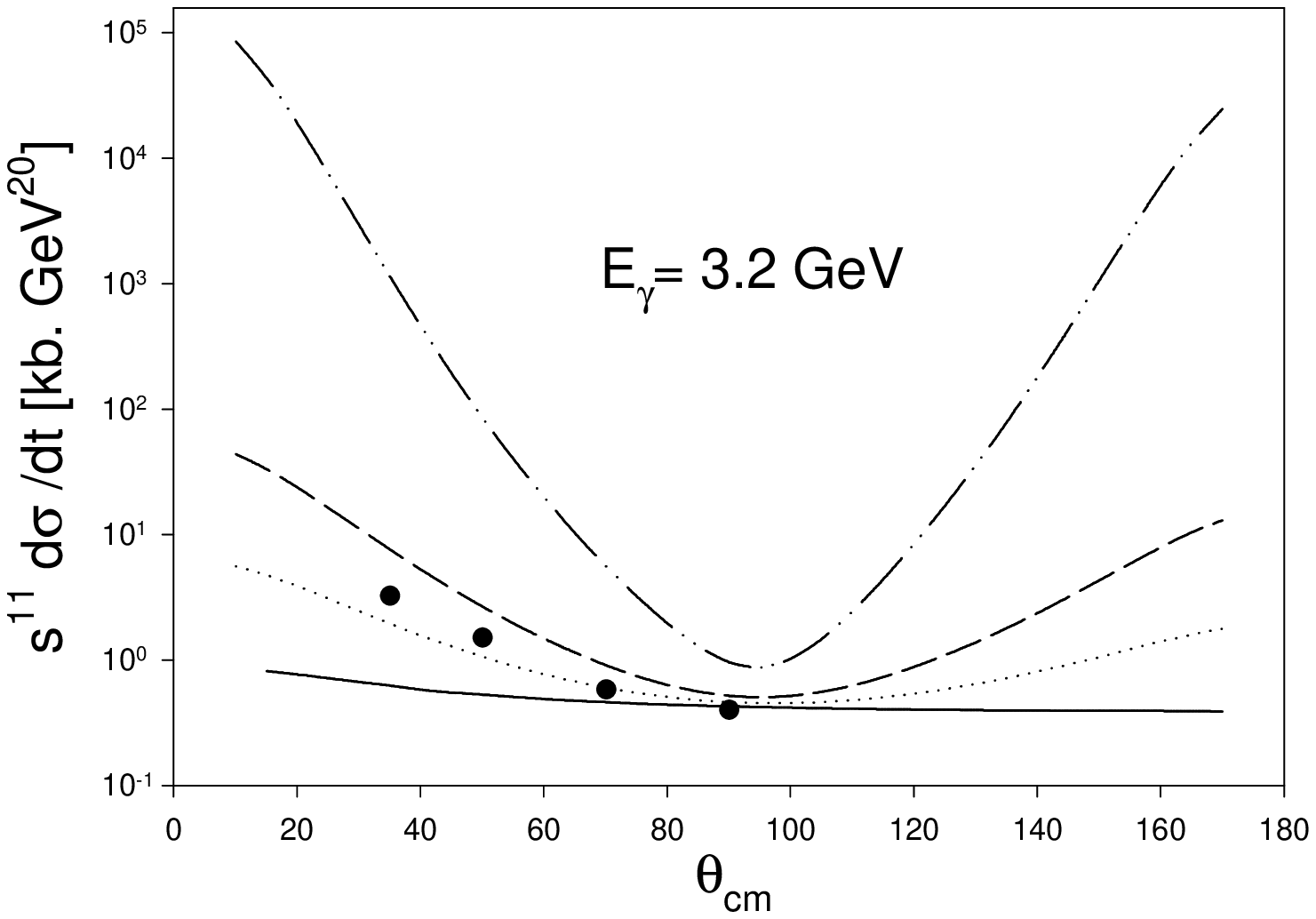}
\newpage
\epsfxsize=18cm
\epsfbox{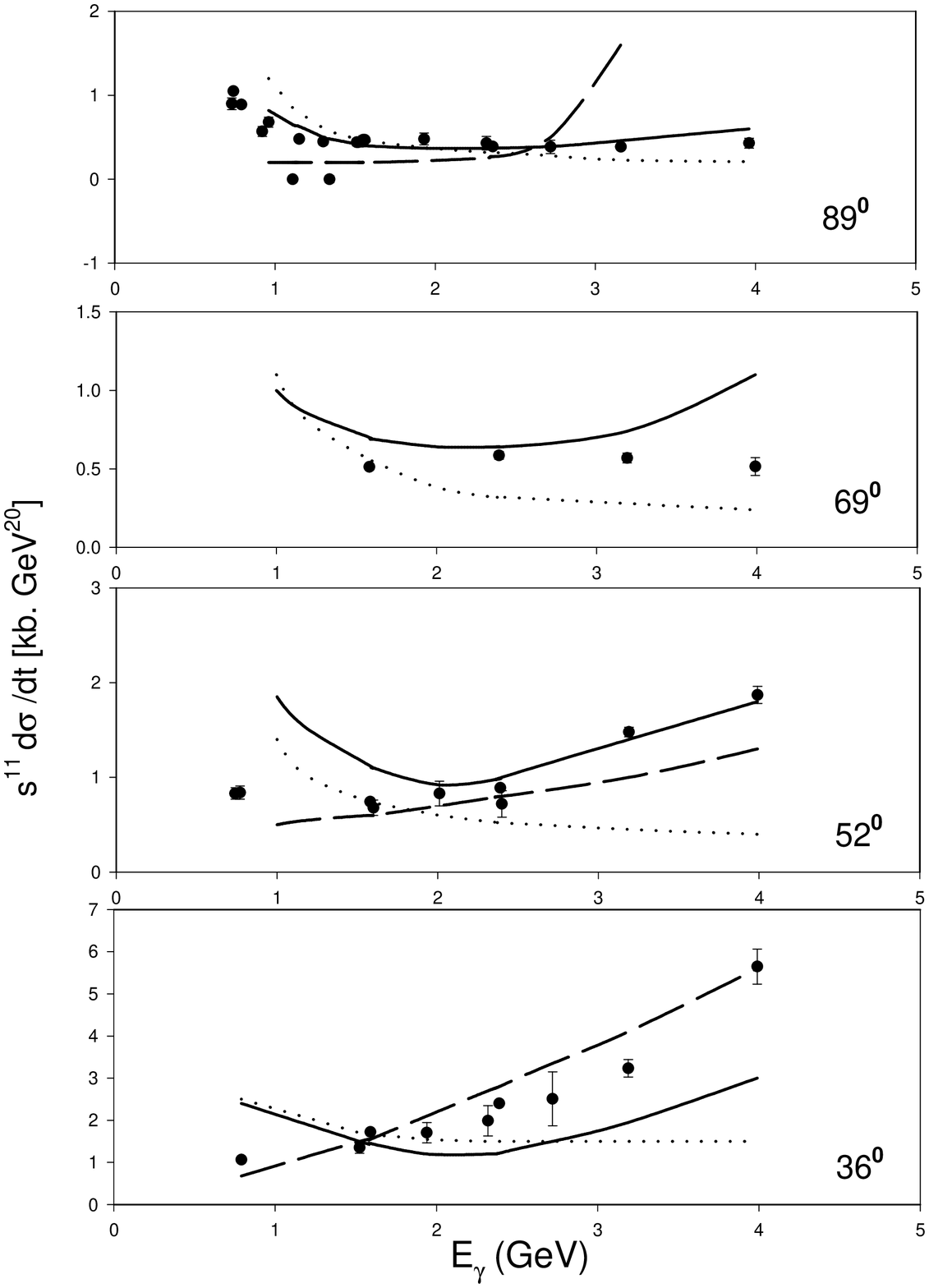}

\end{document}